\documentclass[preprints,article,accept,moreauthors,pdftex]{Definitions/mdpi} 

\usepackage{bm}
\usepackage{bbm}
\usepackage{listings}
\usepackage{subfigure}
\usepackage{amsmath}
\usepackage{amssymb}
\usepackage{booktabs}
\usepackage{hyperref}
\usepackage{erewhon} 
\usepackage[sf,scale=1]{libertine}
\usepackage{fourier}
\usepackage[T1]{fontenc}
\usepackage{algorithm}
\usepackage{algorithmic}

\hypersetup{colorlinks,citecolor=teal}

\firstpage{1} 
\pubvolume{xx}
\issuenum{1}
\articlenumber{1}
\pubyear{2020}
\copyrightyear{2020}
\history{}

\newcommand*\diff{\mathop{}\!\kern0pt\mathrm{d}}

\Title{Entropy of Mersenne-Twisters}
\Author{Fabien Le Floc'h}
\AuthorNames{Fabien Le Floc'h}
\address{}
\corres{Correspondence: F.L.Y.LeFloch@tudelft.nl}
\abstract{The Mersenne-Twister  is one of the most popular generators of uniform pseudo-random numbers. It is used in many numerical libraries and software. In this paper, we look at the Komolgorov entropy of the original Mersenne-Twister, as well as of more modern variations such as the 64-bit Mersenne-Twisters, the Well generators, and the Melg generators.}

\begin{document}
	\section{Introduction}
	The Mersenne-Twister 	\citep{matsumoto1998mersenne}  is one of the most popular generators of uniform pseudo-random numbers. It is used in many numerical libraries and software.
	
	Several variant have been developed over the years, to accommodate some limitations of the original algorithm. 
	
	It has been enhanced to work directly with 64-bit numbers in \citep{nishimura2000tables}. The same paper also describe a variant which increases the number of non-zero terms in the characteristic polynomial. Although rarely adopted in software, the latter has the tremendous advantage to escape much faster from zeroland. Indeed, the original Mersenne-Twister may require more than 700000 numbers to escape from a state with all bits set to zero, except for one. Although a good initialization scheme helps to minimize the issue, the issue may still unexpectedly appear. We give an example of innocuous initialization array, with many ones and zeros.
	
	 \citet{panneton2006improved} propose a more general algorithm, based on the same underlying principles, but using more internal block matrices, while keeping a similar performance profile. This allows for maximal equidistribution properties, a feature not possible in the original design. It also permit the use of non Mersenne prime periods.
	 
	More recently, the algorithm has been slightly modified to obtain a maximally equidistributed variant over 64-bit numbers, by using a lung, but otherwise staying close to the (second) 64-bit Mersenne-Twister variant \citep{harase2018implementing}.
	
 	While introducing his Mixmax random number generator, \citet{savvidy2015mixmax} discusses two properties of the original MT: a low entropy and eigenvalues falling on the unit circle, both of which are detrimental to randomness.
 	
 	In this paper, we start by verifying the claims of Savvidy. We in fact find a larger entropy than stated in \citep{savvidy2015mixmax}. We then compute the eigenvalues and the associated entropy of the different variants of the Mersenne-Twister and discuss their associated properties.
 	
 	\section{Measuring entropy}
In \citep{savvidy2015mixmax}, the Kolmogorov entropy is defined as 
\begin{equation}
	h = -\sum_{k: |\lambda| < 1} \ln |\lambda_k|\,,
\end{equation}
where $(\lambda_k)$ are the eigenvalues of the state transition matrix.

For a k-mixing system,the auto-correlation decay time $\tau_0$ is related to the entropy as
 $\tau_0 \leq 1 / h$. It is thus desirable that most of the eigenvalues of the
matrix should lie as far as possible away from it. Without involving the theory around Komolgorov entropy, it is intuitive that randomness will increase as the eigenvalues of the linear system spread out.

An $\mathbb{F}_2$-linear generator may be defined though its state transition matrix by the relation \citep{panneton2006improved}
\begin{align*}
	\bm{x}_i &= 	\bm{B} \bm{x}_{i-1} \mod 2\,,\\
	\bm{y}_i &= \bm{C} \bm{x}_i \mod 2\,.
\end{align*} 
where $\bm{x}_i = (x_{i,k-1},...,x_{i,0})^T  \in \mathbb{F}^k_2$ and $\bm{y}_i = (y_{i,w-1}, ..., y_{i,0})^T \in \mathbb{F}^w_2$ are respectively the $k$-bit
state and the $w$-bit output vector at step $i$, $\bm{B}$ and $\bm{C}$ are a $w \times k$ transition matrix
and a $w \times k$ output transformation matrix with elements in $\mathbb{F}_2$, $k$ and $w$ are
positive integers. 

In the Mersenne-twister, the output transformation takes the first word $(x_{i,k-1},..,x_{i,k-1-w})$ of the state, and eventually tempers it to improve equidistribution properties. 

For the original Mersenne-twister MT19937 \citep{matsumoto1998mersenne}, the matrix $B$ of size $(nw-r \times nw-r)$ is expressed in terms of word vectors $\bm{x}_j^w$ (instead of bits) as follows
\begin{equation}
\bm{x}_i	= (\bm{x}_{i,n-1}^w, \bm{x}_{i,n-2}^w, ..., \overline{\bm{x}}_{i,0}^w) = 	(\bm{x}_{i-1,n-1}^w, \bm{x}_{i-1,n-2}^w, ..., \overline{\bm{x}}_{i-1,0}^w) \bm{B}^T
\end{equation} 
where $\overline{\bm{x}}^w_{i,j}$,  represents the $w-r$ most significant $\bm{x}^w_{i,j}$. The matrix $B$ reads,
\begin{equation*}
	\bm{B}^T = \begin{pmatrix}
		& \bm{I}_w & & &\\
		&     &\bm{I}_w & &\\
		\bm{I}_w & & & \ddots&\\
		& & & &\bm{I}_{w-r} \\
		\bm{S} & & & &\\
	\end{pmatrix}\,,
\end{equation*}
with $n=624, m=397, w=32, r=31$, and
\begin{equation*}
	\bm{S} = \begin{pmatrix}
		0 & \bm{I}_r\\
		\bm{I}_{w-r} & 0
	\end{pmatrix} \bm{A}\,.
\end{equation*}
The matrix $A$ is defined as
\begin{equation*}
	\bm{A} = \begin{pmatrix}
		& 1 &   & & \\
		&   & 1 & & \\
		&   &   &\ddots & \\
		&   &   &      & 1\\
		a_{w-1} & a_{w-2} & \dots & \dots & a_0
	\end{pmatrix}\,,
\end{equation*}
with $\bm{a} = (a_{w-1},a_{w-2},...,a_0) = \texttt{0x9908B0DF}$.

Thus, in the MT19937 generator, we have \begin{equation}
	\bm{x}_{i,j}^w = \bm{x}_{i-1,j+1}^w \quad \textmd{ for } \quad j=n-2,...,0\,.\label{eqn:consecutive}
\end{equation} 
This property is however not true in general for the other variants.

The transition matrix defines the iteration, and is thus the relevant one for the study of entropy.
For $\mathbb{F}_2$-linear generators, the eigenvalues of the transition matrix correspond to the roots of its characteristic polynomial in $\mathbb{Z}$. In particular, those are different from the eigenvalues of the characteristic polynomial commonly used for skipping ahead, or verifying primitivity, which is stated in $\mathbb{F}_2$. For the original Mersenne-Twister, we give the polynomial in  $\mathbb{Z}$ in Appendix \ref{appendix:mt_char}.

In practice however, we will not use the characteristic polynomial but work directly from the state transition matrix.
For an $\mathbb{F}_2$-linear generator, the matrix can be recovered by seeding the standard basis successively, and fetching the state of the generator after one iteration as described in Algorithm \ref{alg:state}.

\begin{algorithm}
\linespread{1.35}\selectfont
	\caption{Print the state transition matrix for Mersenne-Twister generators \label{alg:state}}
	\begin{algorithmic}[1]
		\STATE $i \leftarrow n*w-1$ 
		\WHILE{$i \geq 0$}		
		\STATE $indicator$ is an array of $n$ integers, each of $w$-bit size, initialized to zero
		\STATE $k \leftarrow i/w$
		\STATE $indicator[k] \leftarrow 1 << (i \mod w)$ \COMMENT{set the $i$-th bit to one}
		\STATE initialize the random number generator with the state $indicator$
		\STATE compute the next random number
		\FOR{$k \leftarrow 0$ to $n-1$} 
		\STATE $indicator[k] \leftarrow state[k+1 \mod n]$		\COMMENT{copy the new state to the array $indicator$ with the correct indexing}
		\ENDFOR
		\FOR{$k \leftarrow n-1$ to $0$, $j \leftarrow w-1$ to $0$} 
				\IF{$indicator[k] \&  (1 <<  j) \neq 0$} 
					\PRINT 1
				\ELSE 
				   \PRINT 0
	  			\ENDIF
		\ENDFOR
		\PRINT new line
		\ENDWHILE
	\end{algorithmic}
\end{algorithm}

For the Melg generators of \citet{harase2018implementing}, the lung is always the last element of the state and does not move around.
For the Well generators, the convention of the state matrix described in \cite{panneton2006improved} is to use the most significant bits of each word first. The algorithm is reversed, from $I=0$ to $N*w-1$ (lines 1 and 2), $K$ and $J$ are reversed in similar fashion. This leads to Algorithm \ref{alg:state_well} in Appendix \ref{appendix:algo}.

Working directly with the state transition matrix avoids the numerical issues of the characteristic polynomial approach, which arise because of the large degrees involved.

\section{Results}
\begin{figure}[h]
	\centering{
	\includegraphics[width=.45\textwidth]{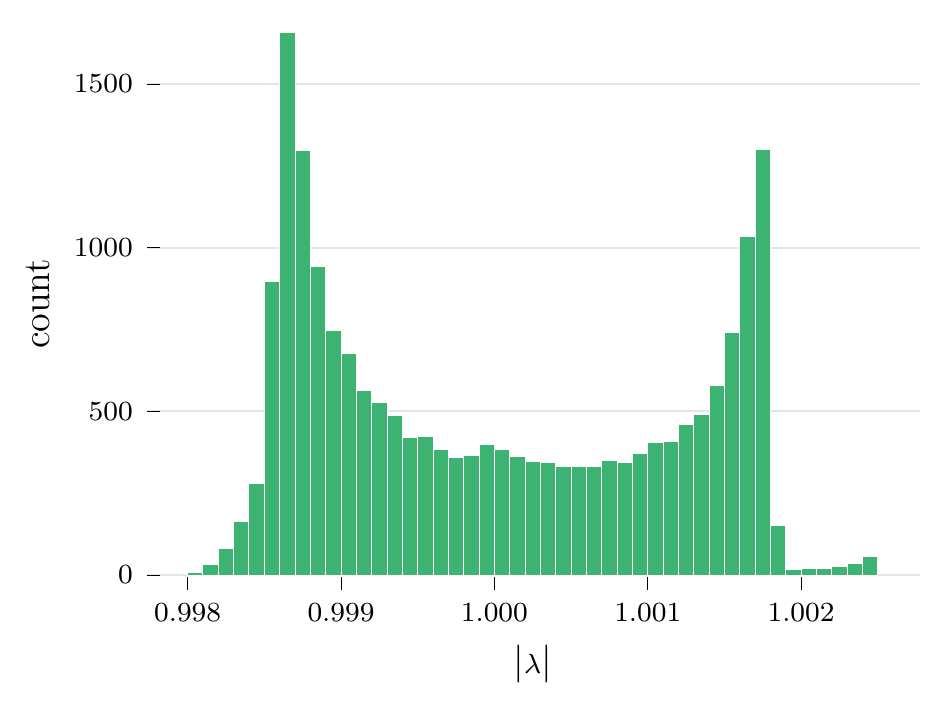}
			\hfill
			\includegraphics[width=.45\textwidth]{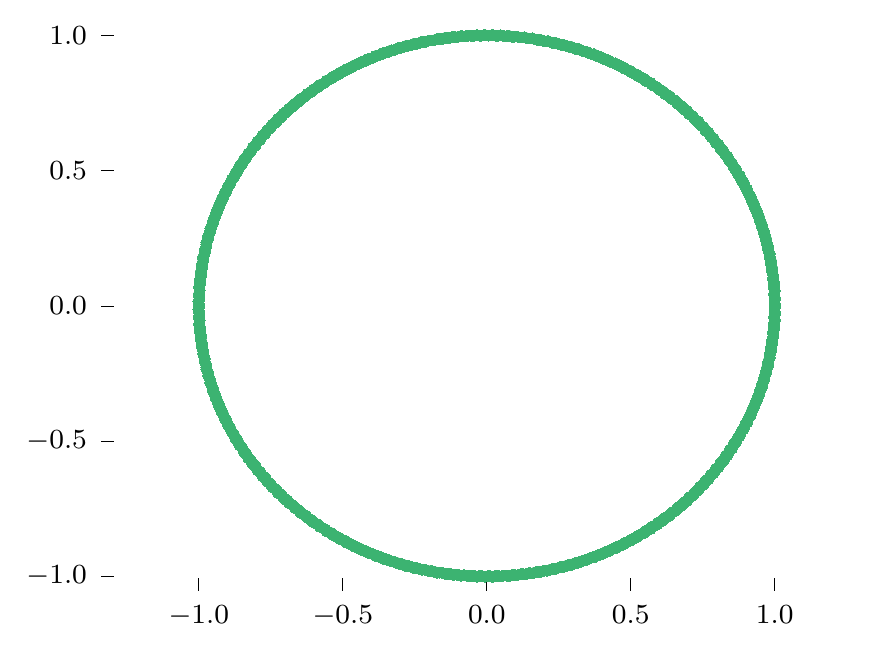} 
		}
		\caption{Eigenvalues of the transition matrix for the 32-bit Mersenne-Twister: histogram (left) and complex plane representation (right).\label{fig:mt19937}}
\end{figure}

	The eigenvalues of the original version of Mersenne-Twister, with a period of $2^{19937}-1$, all fall close to unit circle (Figure \ref{fig:mt19937}). The modulus of the smallest eigenvalue is 0.99806, and the modulus of the largest eigenvalue is 1.00246. While \citet{savvidy2015mixmax} finds an entropy of 4.8, we measure an entropy of 10.49. We double checked our results by solving the roots of the characteristic polynomial with a robust polynomial roots solver \citep{aurentz2018fast} and quadruple precision arithmetic.	
	
If we look at the eigenvalues of other $\mathbb{F}_2$-linear generators, we notice that for a very large period of $2^{19937}-1$, and thus a characteristic polynomial degree of 19937, the eigenvalues always tend to be very close to the circle (see Figure \ref{fig:eig19937} in Appendix \ref{appendix:eigenvalues}), even though the entropy increases substantially for some and may be close to the entropy of the generators with a smaller period. 
For example, the ID3 version of the 64-bit Mersenne-Twister (named MT19937x64ID3 thereafter) has twice the entropy of the ID1 version of the 64-bit Mersenne-Twister. With a large characteristic polynomial degree, the eigenvalues are likely to be very close to the unit circle: for a value strongly greater than one, the polynomial value explodes because of the dominant term of degree 19937. 

\begin{figure}[h]
	\centering{
		\includegraphics[width=.45\textwidth]{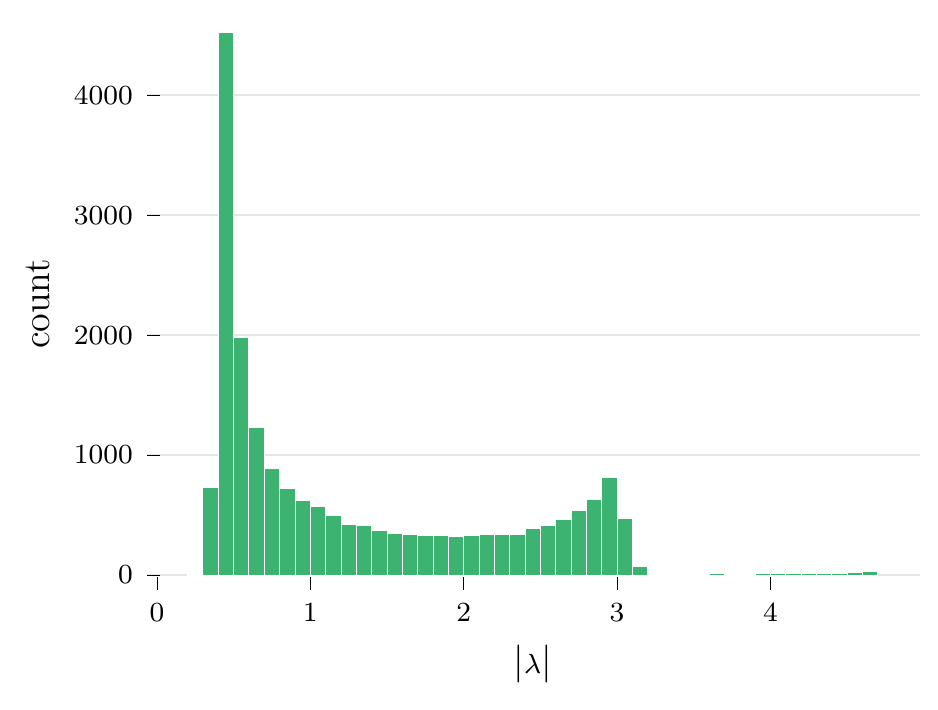}
		\hfill
		\includegraphics[width=.45\textwidth]{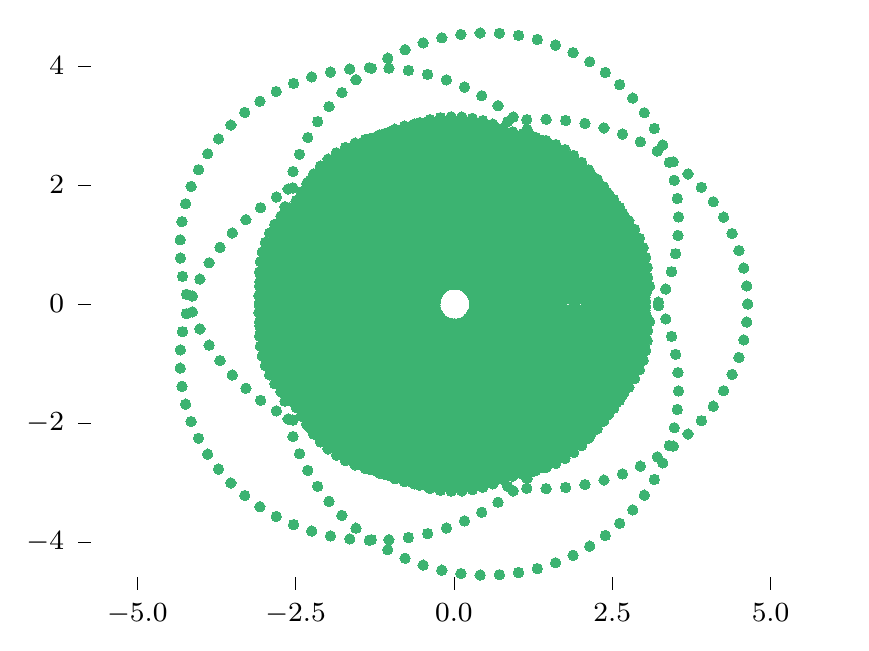} 
	}
	\caption{Eigenvalues of the transition matrix to the power 624 for the 32-bit Mersenne-Twister: histogram (left) and complex plane representation (right).\label{fig:mt19937_624}}
\end{figure}

In reality, only 32 bits (or 64 bits for the 64-bit generators) are produced by one application of the state transition matrix $B$. In order to be comparable to the Mixmax generator of \citet{savvidy2015mixmax}, one would need to consider $n$ applications of the matrix, where $n=624$ for the 32-bit Mersenne-Twister as the state of the Mersenne-Twister is composed of 624 words. And then, we would see much larger eigenvalues, since the eigenvalues of $B^n$ are simply the $n$-th power of the eigenvalues of $B$ (Figure \ref{fig:mt19937_624}). For the 32-bit and 64-bit Mersenne-Twisters, the $n$-th power is not merely the matrix to skip to the $n$-th point in the sequence, because the $n-1$ previous numbers $x_{n-1,1}^w,...,x_{1,1}^w$ are also in the output vector, as consequence of Equation \ref{eqn:consecutive}. This is however not true any more for the Melg and Well generators, we believe however that the eigenvalues for the $n$-th power are still representative of the entropy of the $n$-dimensional output.

 For the same reason, $\mathbb{F}_2$-linear generators working on 64-bit words have twice the entropy of the ones working on 32-bit words. If we want to compare the two kinds of generator, a more relevant measure is the entropy per bit of random number generated (of size one word).

From Table \ref{tbl:entropy}, we notice that the entropy per bit is the highest for the Well generators. In particular, Well607b has the largest entropy per bit.	
	\begin{table}[h]
		\caption{Entropy of the different random number generators. The performance is normalized towards the 32-bit Mersenne-Twister MT19937. MT19937 took 3.8ms to output one million double floating point numbers on an AMD Ryzen 1700 processor running OpenJDK 15 under Linux. \label{tbl:entropy}}
		\centering{
			\begin{tabular}{lrrrrr}\toprule
				Generator &word size $w$ & Entropy & Entropy per bit & $N_1$ & Performance\\\midrule
				MT19937 & 32 & 10.49 & 0.32 & 135 & 1.00\\
				MT19937-64ID1 & 64 & 20.75&  0.32& 285 & 0.87\\
				MT19937-64ID3 & 64 &40.05 & 0.63 & 5795 & 0.49\\
				Well19937a & 32 & 23.58 & 0.74 & 8585 & 0.68\\
				Well1024a & 32 &25.79& 0.81 &407 & 1.06\\
				\bf{Well607b} & 32 &26.66 & \bf{0.83} & 313 & 0.83\\
				Melg607 & 64 &42.83 & 0.67 &313 & 0.85\\
				Melg19937 & 64 &41.63 & 0.65 & 9603 & 0.85\\
				\bottomrule
		\end{tabular}}
	\end{table}
There seems to be a direct relationship with the blocks in the first block column of the transition matrix and the entropy. The MT19937 has two non-zero blocks of dimension $w \times w$. The 64-bit version MT19937x64ID1 has the same configuration, but where $w$ is twice the length, and this produces approximately twice the entropy. The version MT19937x64ID3 has four non-zero blocks (so another factor two), and leads to four times this entropy. Similarly, the Melg generators have four non-zero blocks, and the Well generators have four to five non-zero blocks. We 	notice a linear relationship between the entropy and the number of non-zero blocks in the first block-column (also taking there size into account).

\begin{figure}[h]
	\centering{
		\includegraphics[width=.45\textwidth]{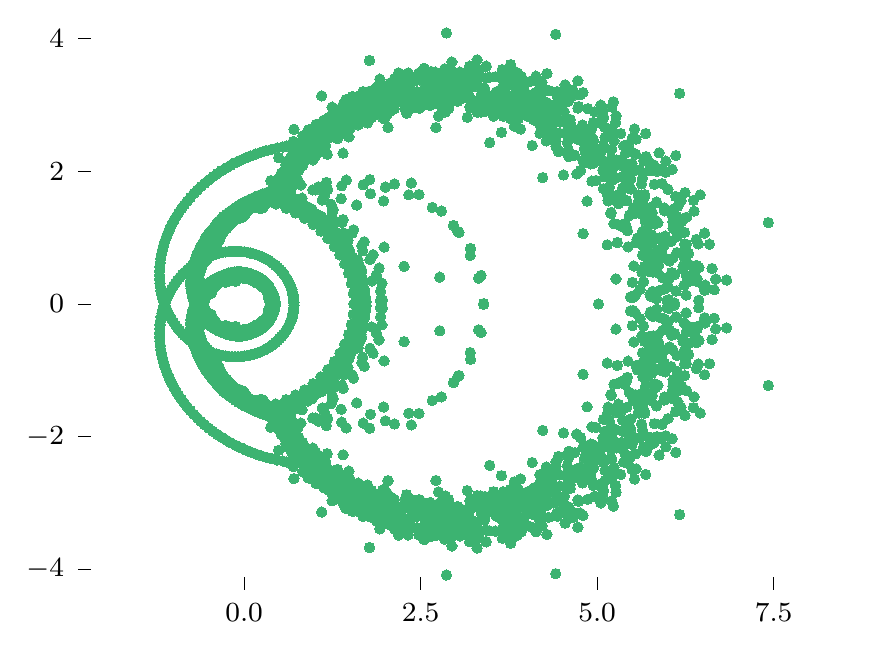}
		\hfill
		\includegraphics[width=.45\textwidth]{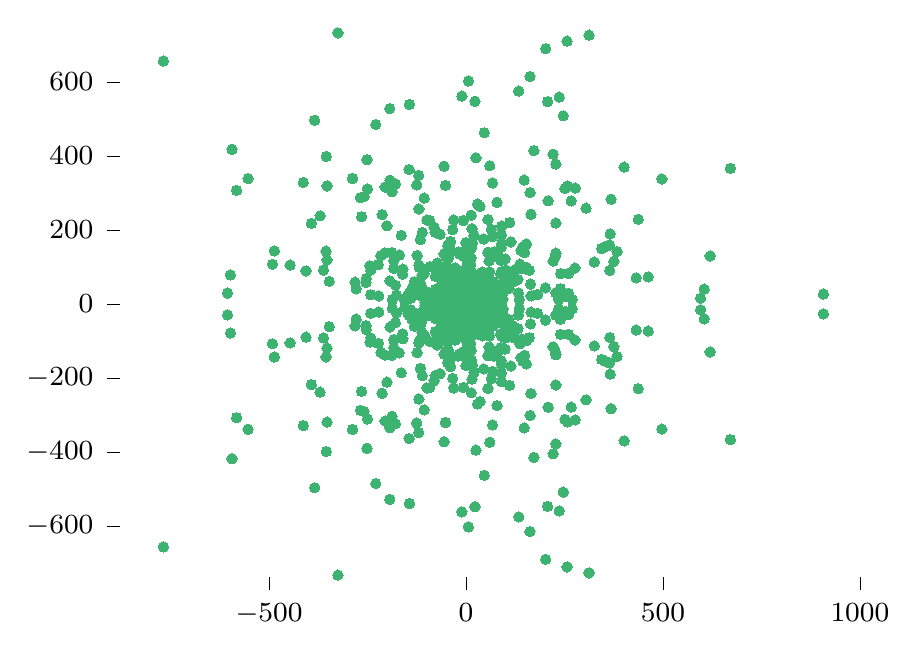} 
	}
	\caption{Eigenvalues of the transition matrix to raised to the power 312 for the 64-bit Mersenne-Twisters: ID1 (left) and ID3 (right).\label{fig:mt19937x64_312}}
\end{figure}

On Figure \ref{fig:mt19937x64_312}, we look at the state matrix, raised to the power 312, for the 64-bit Mersenne-Twisters MT19937x64ID1 and MT19937x64ID3. It corresponds to a full update of the state vector.
 As expected, the larger entropy translates to a pronounced increase of the spread between eigenvalues. The axis scale is multiplied by a factor larger than two hundred between the ID1 and the ID3 versions.

Regarding performance, the Well1024a is the fastest at generating one million random numbers in Java, closely followed by the 32-bit MT19937. The MT19937-64ID3 is the slowest, and the Melg generators are better performing alternative 64-bit generators, with good entropy properties.

\section{Reaching and escaping zeroland}
Following the methodology of \citet{panneton2006improved}, we consider the initial states filled with zero bits, except for one bit, and measure the number iterations it takes to reach a state with a balanced number of zero and one bits. There are $k$ such states, and we we let $y_i^{(j)}$ be the output vector $y_i$ at iteration $i$ for a given initial state $\bm{x}_0 = \bm{e}_j$ where $\bm{e}_j$ is the $j$-th unit vector, for $j=1,...,k$. Let $H(\bm{x})$ be the Hamming weight of a bit vector $\bm{x}$, i.e., the number of bits set to 1, the moving average $\gamma_{n,p}$ is defined by
\begin{equation}
	\gamma_{n,p} = \frac{1}{pkw} \sum_{i=n}^{n+p-1} \sum_{j=1}^k H\left(\bm{y}_i^{(j)}\right)\,.
\end{equation}
Under the null hypothesis, $\gamma_{n,p}$ should be approximately normally distributed with mean $1/2$ and variance $1/(4pkw)$.

\begin{figure}[h]
	\centering{
		\subfigure[\label{fig:hamming_19937_av}$\gamma_{n,100}$, starting from unit vectors ($k=19937$)]{		\includegraphics[width=.45\textwidth]{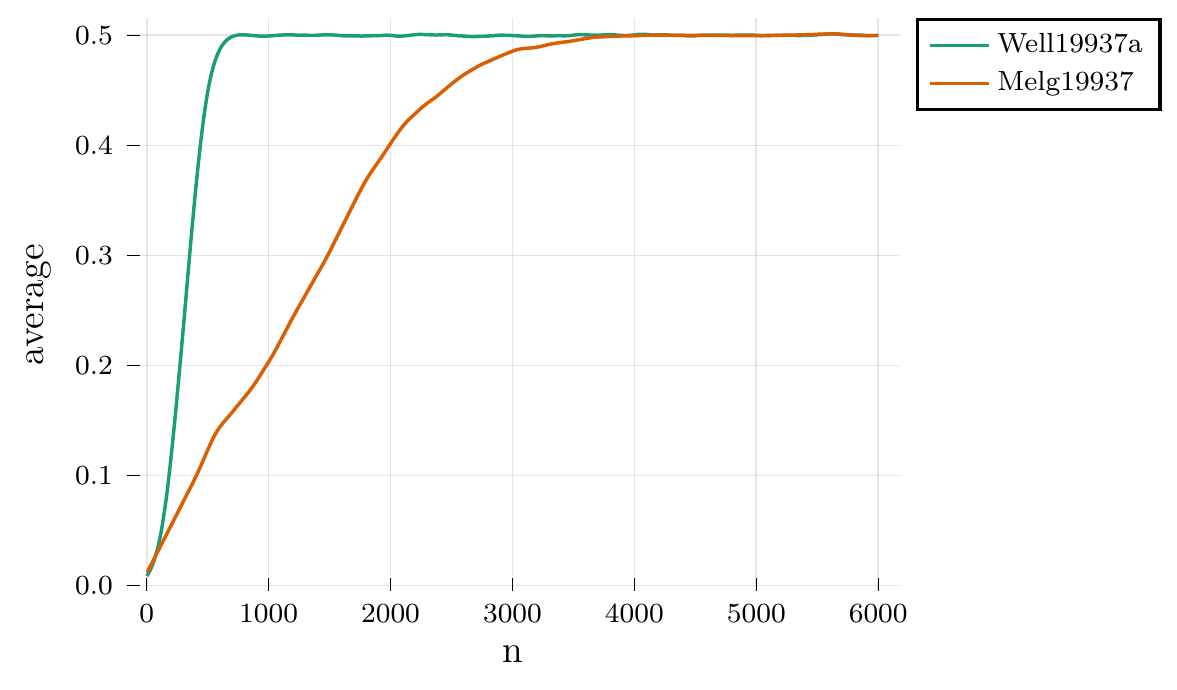}
		}
		\hfill
			\subfigure[\label{fig:badseed_19937}$\gamma_{n,624}$, starting from a bad seed ($k=1$)]{	
		\includegraphics[width=.45\textwidth]{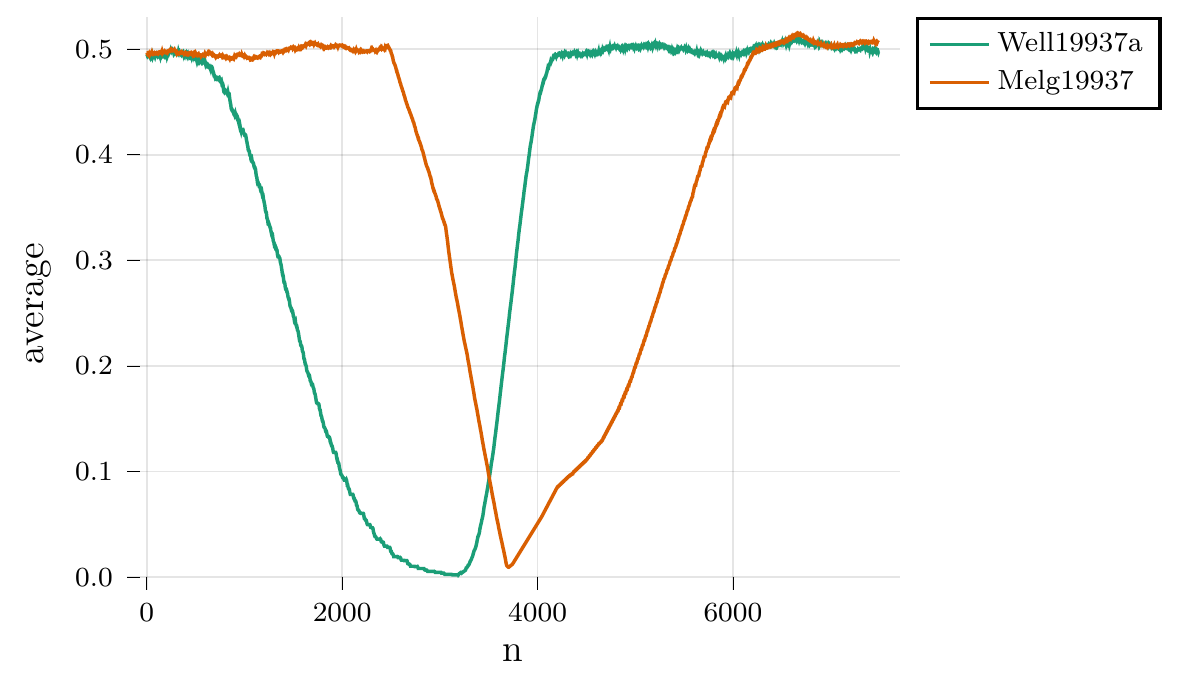}
	} 
	}
	\caption{Hamming weight after $n$ iterations.\label{fig:hamming}}
\end{figure}

On Figure \ref{fig:hamming_19937_av}, we plot $\gamma_{n,100}$ as a function of $n$. In order to compare 32-bit with 64-bit generators, we normalized the number of iterations $n$ according to the size of one word: one iteration of a 64-bit generator counts for two iterations of a 32-bit generator. Similarly $p$ is halved for 64-bit generators.
Well19937a requires around 730 iterations to reach a balanced state, while Melg19937 requires $2\times2040$ iterations, even though its characteristic polynomial has a larger number of ones ($N_1$ in Table \ref{tbl:entropy}). Even without normalization, the Melg generator requires more iterations than the Well generator. The measure of entropy per bit is however consistent with this result: it it higher for Well19937a than for Melg19937.

Of course, in practice, the random number generators will not be used with such a state. A good seed initialization procedure, such as the one used in the Melg generators, provides a simple remedy. The reference implementations of Well generators do not include such an initialization procedure, and this sometimes leads to artificially bad results in papers, which study the behaviour of random number generators with close seeds such as \citep{bhattacharjee2018search}. A real world implementation will however implement a good initialization procedure.

With a good initialization, even if the probability is then very low, it is never fully guaranteed that, accidentally, some seed will lead to a unfavorable state. We give an example of an random looking seed, which could well be the produce of a good initialization procedure as an illustration in Appendix \ref{appendix:badseed}. In order to find this special seed, we simply jump-ahead close to the end of the period, starting with a seed composed of one non-zero bit. Figure \ref{fig:badseed_19937} shows an example of special seed, where $\gamma_{n,624}$ nearly reaches zero after 3500 iterations. There is less difference between the Well and the Melg generators on this particular example, but this could be because we only look at a specific seed, instead of an average over many seeds.

The equidistribution properties of the generators guarantee that those awkward states, with a very large percentage of zeros, are present in a period. In order to find a good initial seed, a good alternative may be to start with a seed composed of one non-zero bit, and jump-ahead sufficiently far away, but not too far. Then we may obtain a good starting point as many of the bad states are in the relatively recent past.

The issue is nearly invisible for generators with a smaller state such as Well607b: they escape zeroland after less than 50 iterations: too quickly to matter statistically. 

\section{Conclusion}
The entropy reveals some properties of the $\mathbb{F}_2$ linear generators. In a similar fashion as the number of non-zero bits in the characteristic polynomial, a low entropy is related to a low mixing of bits. Entropy however reveals more about the underlying generator as it seems linearly linked to the number of non-zero blocks in the first block row of the transition matrix. 

The Well generators have the largest entropy per bit among the studied generators.

\funding{This research received no external funding.}
\conflictsofinterest{The authors declare no conflict of interest.}
\acknowledgments{Fruitful conversations with K. Savvidy.}
\externalbibliography{yes}
\bibliography{well64.bib}
\appendixtitles{no}

	\appendix

\section{Twisted GFSR characteristic polynomial}\label{appendix:gfsr_char}
We consider the $(nw \times nw)$-matrix $B$ defined as 
\begin{equation*}
	B = \begin{pmatrix}
		& I_w & & &\\
		&     &I_w & &\\
		I_w & & & \ddots&\\
		& & & &I_{w} \\
		S & & & &\\
	\end{pmatrix}\,,
\end{equation*}
where, $I_w$ is the identity matrix of size $w$, in the first block column, the first row of $I_w$ is at the row position $((n-1-m)w+1)$ in $B$.

In \citep[Appendix A.1]{matsumoto1992twisted}, it is proven that, for the case $r=0$, the characteristic polynomial $\phi_B(t) = \det(t I_{nw} - B)$ of $B$ is
\begin{equation*}
	\phi_B(t) = \phi_S (t^n + t^m)\,,
\end{equation*}
where $\phi_S$ is the characteristic polynomial of the $(w,w)$-matrix $S$.

This is true in $\mathbb{F}_2$ but not in the general case. The correct formula is
\begin{equation}
	\phi_B(t) = \phi_S (t^n - t^m)\,,
\end{equation}
\begin{proof}
	We have 
	\begin{equation*}
		\phi_B(t) = \det(t I_{nw-r} - B) = \det \begin{pmatrix}
			t I_w& -I_w & & &\\
			&t I_w     &- I_w & &\\
			-I_w & &t I_w & \ddots&\\
			& & &t I_{w-r} &-I_{w-r} \\
			-S & & & &t I_w\\
		\end{pmatrix}\,.
	\end{equation*}
	The determinant stays unchanged by linear combinations of columns. We add second block column times $t$ to first column to obtain
	\begin{equation*}
		\phi_B(t) = \det \begin{pmatrix}
			0      & -I_w & & & &\\
			t^2 I_w&t I_w  & - I_w & & &\\
			\vdots & &t I_w & &\\
			-I_w   & & &\ddots & \ddots&\\
			&   & &&t I_{w}  &-I_{w} \\
			-S     & & & & &t I_w\\
		\end{pmatrix}\,.
	\end{equation*}
	Then then we continue with $t^k$ times $(k+1)$-th block column, until line $(n-1-m)w+1$, we obtain
	\begin{equation*}
		\phi_B(t) = \det \begin{pmatrix}
			0      & -I_w & & & &\\
			0 &t I_w  & - I_w & & &\\
			\vdots & &t I_w & &\\
			t^{n-m} I_w -I_w   & & &\ddots & \ddots&\\
			&   & &&t I_{w-r}  &-I_{w} \\
			-S     & & & & &t I_w\\
		\end{pmatrix}\,.
	\end{equation*}
	We now subtract $t^k(t^{n-m} - 1)$ times the $(n-m)+k$-th block column, until line $(n-1)w+1$:
	\begin{equation*}
		\phi_B(t) = \det \begin{pmatrix}
			0      & -I_w & & & &\\
			0 &t I_w  & - I_w & & &\\
			\vdots & &t I_w & &\\
			& & &\ddots & \ddots&\\
			0 &   & &&t I_{w-r}  &-I_{w-r} \\
			t^{m}(t^{n-m} -1)I_{w}-S     & & & & &t I_w\\
		\end{pmatrix}\,.
	\end{equation*}
	Now we reduce the $j$-th block column, by adding $t^k$ times the $k+j$ block, for $j=2,...,n-1$ and $k=1,...,n-j$. This leads to
	\begin{equation*}
		\phi_B(t) = \det \begin{pmatrix}
			0      & -I_w & & & &\\
			0 & 0  & - I_w & & &\\
			\vdots & &0 & &\\
			& & &\ddots & \ddots&\\
			0 &   & &&0  &-I_{w-r} \\
			t^{m}(t^{n-m} -1)I_{w}-S     &  t^{n-1}I_w & & & t^2 I_w &t I_w\\
		\end{pmatrix}\,.
	\end{equation*}
	We use $(n-1)$ permutations to move the first block column to the last position, we end up with a block-triangular matrix, the determinant is the product of the diagonal determinants.
	\begin{equation*}
		\phi_B(t) = \det(	t^{m}(t^{n-m} -1)I_{w}-S ) = \phi_S(t^n -t^m)\,.
	\end{equation*}
\end{proof}	

\section{Mersenne-Twister characteristic polynomial}\label{appendix:mt_char}
In the Mersenne-twister random number generator MT19937 \citep{matsumoto1998mersenne}, the matrix $B$ corresponds to the state transition matrix, and, compared to the GFSR of Appendix \ref{appendix:gfsr_char}, is slightly modified to accommodate for a size $(nw-r \times nw-r)$. It is defined as follows:
\begin{equation*}
	B = \begin{pmatrix}
		& I_w & & &\\
		&     &I_w & &\\
		I_w & & & \ddots&\\
		& & & &I_{w-r} \\
		S & & & &\\
	\end{pmatrix}\,,
\end{equation*}
with $n=624, m=397, w=32, r=31$, and the matrix $S$ reads
\begin{equation*}
	S = \begin{pmatrix}
		0 & I_r\\
		I_{w-r} & 0
	\end{pmatrix} A\,.
\end{equation*}
The matrix $A$ is defined as
\begin{equation*}
	A = \begin{pmatrix}
		& 1 &   & & \\
		&   & 1 & & \\
		&   &   &\ddots & \\
		&   &   &      & 1\\
		a_{w-1} & a_{w-2} & \dots & \dots & a_0
	\end{pmatrix}\,,
\end{equation*}
with $a = (a_{w-1},a_{w-2},...,a_0) = \texttt{0x9908B0DF}$.
In particular, $\phi_A(t) = t^w - \sum_{i=0}^{w-1} a_{i} t^{w-i-1}$. 

For $w-r = 1$, the $S$ matrix reads
\begin{equation*}
	S = \begin{pmatrix}
		0	& 0  & 1 & & \\
		&   &   &\ddots & \\
		&   &   &      & 1\\
		a_{w-1} & a_{w-2} & \dots & \dots & a_0\\
		0	& 1 &  0 & \dots & 0
	\end{pmatrix}\,,
\end{equation*}
With similar column transformations as for the twisted GFSR generator (see Appendix \ref{appendix:gfsr_char}), the characteristic polynomial for the Mersenne-Twister reads
\begin{equation}
	\phi_B(t) = 	(t^n-t^m)^{w-r}(t^{n-1}-t^{m-1})^r - (t^n-t^m)^{w-r}\sum_{i=0}^{r-2} a_i (t^{n-1}-t^{m-1})^{r-1-i}- \sum_{i=r-1}^{w-1} a_i (t^n-t^m)^{w-i-1}\,.
\end{equation}
For the MT19937, we have in particular,
\begin{align*}
	\phi_B(t) &=
	(t^n-t^m)(t^{n-1}-t^{m-1})^{31}-(t^n-t^m)(t^{n-1}-t^{m-1})^{30}-(t^n-t^m)(t^{n-1}-t^{m-1})^{29}-(t^n-t^m)(t^{n-1}-t^{m-1})^{28}\\
	&-(t^n-t^m)(t^{n-1}-t^{m-1})^{27}-(t^n-t^m)(t^{n-1}-t^{m-1})^{26}-(t^n-t^m)(t^{n-1}-t^{m-1})^{24}-(t^n-t^m)(t^{n-1}-t^{m-1})^{23}\\
	&-(t^n-t^m)(t^{n-1}-t^{m-1})^{18}-(t^n-t^m)(t^{n-1}-t^{m-1})^{17}-(t^n-t^m)(t^{n-1}-t^{m-1})^{15}-(t^n-t^m)(t^{n-1}-t^{m-1})^{11}\\
	&-(t^n-t^m)(t^{n-1}-t^{m-1})^6-(t^n-t^m)(t^{n-1}-t^{m-1})^3-(t^n-t^m)(t^{n-1}-t^{m-1})^2-1\,.
\end{align*}

\section{Algorithm for state transition matrix}\label{appendix:algo}

\begin{algorithm}
	\linespread{1.35}\selectfont
	\caption{Print the state transition matrix for Well generators \label{alg:state_well}}
	\begin{algorithmic}[1]
		\STATE $I \leftarrow 0$ 
		\WHILE{$I \geq N*w-1$}		
		\STATE $indicator$ is an array of $N$ $w$-bit integers 		
		\STATE $K \leftarrow I/w$
		\STATE $indicator[K] \leftarrow 1 << (w-1-i \mod w)$ 
		\STATE initialize the random number generator with the state $indicator$
		\STATE compute the next random number
		\FOR{$K \leftarrow 0$ to $N-1$} 
		\STATE $indicator[K] \leftarrow state[K+N-1 \mod N]$		\COMMENT{copy the new state to the array $indicator$ with the correct indexing}
		\ENDFOR
		\FOR{$K \leftarrow 0$ to $N-1$, $J \leftarrow 0$ to $W-1$} 
		\IF{$indicator[K] \&  (1 << w-1-j) \neq 0$} 
		\PRINT 1
		\ELSE 
		\PRINT 0
		\ENDIF
		\ENDFOR
		\PRINT new line
		\ENDWHILE
	\end{algorithmic}
\end{algorithm}

\section{Bad seeds}\label{appendix:badseed}
\subsection{Melg19937}
\tiny{
	\begin{verbatim}
seed = { 0xedf329eaf017de19L, 0xbcd59767baffa8a4L, 0x607fab7caa6557a3L, 0x51836b8de6aee5dfL,
	0x83fd89742e80cf41L, 0x46b9f946143ddf01L, 0x1fc0637bd4a0995cL, 0xb9a96f9a6b97c574L, 0xd8e318b4125e1d6aL,
	0x2d26bd8d547affa6L, 0x88b5ea4a671d1e4dL, 0xfbde1f5255bccb1bL, 0xb0cb8135b58ffe1cL, 0x2de84bfe5a28d517L,
	0xce6799e73f58b261L, 0x981e2e620ced06dL, 0x424c7c385f53778eL, 0xf8db29aeb1e26a63L, 0x41c7970a76dffd52L,
	0x5d5d6ca8f7fed43aL, 0x8f5fec07ad6ecb4L, 0x1aa208d349b79b15L, 0x2635a1a56753541L, 0x4f6cc6109f3906c7L,
	0xe3b24037fdc0b929L, 0x991a4d1dc5398cfeL, 0x8b7a408f3057502dL, 0xf7edfc08faaecf2cL, 0x574ad9b5649e9460L,
	0x41149d7f1bb48e10L, 0x2450c510c8bb7acL, 0x38d6f51518ead4d7L, 0x35ec0e2e37289f83L, 0xb87f27714bed7f99L,
	0xf9d779eeee811ebeL, 0x891f83398ae69609L, 0xa1df41f525874e23L, 0xe422baaacb9267b8L, 0x3bb16e9e6929efc8L,
	0x989881f1860eb1f6L, 0xa5424b009e0232b0L, 0x5c3159cbfb5e4d00L, 0xc07765e46ade9639L, 0xc4eaa62837cf66aL,
	0xe68f977f2d38606cL, 0x9f8959695ce1dd91L, 0xcdbf7a8df7e7a40eL, 0xe3fe689ea7281711L, 0x8552651ee3c64347L,
	0x9c9436026c40528bL, 0x7c8c7436667af2c7L, 0xe9cad89cad1ea3e7L, 0xe4ce8283b8beff0aL, 0xed67584cf3e4ad01L,
	0xfa4f0d64b9c63e2cL, 0xc7604dc55703895cL, 0x89a4f7b465b8ff9bL, 0x2c2fb1688f931701L, 0x1507d5bacea15bd3L,
	0xb36181d2f39d4deaL, 0xa89ac6f429a257d3L, 0x83db5f5a132e96d0L, 0xa960a59ea65bfe1aL, 0xbb30bdbfa19c04c1L,
	0x3dd1e4dc3a2abc43L, 0xd160f0b1181a2385L, 0xdd253a8dfeb038e4L, 0x9cdde1a8c527833fL, 0x2f98b9ad4faeb385L,
	0x9463f8d2e2d3d15cL, 0x5a7e4e3a2dc2a6c3L, 0x1715e353342515d0L, 0xa1f77ede47509d6L, 0x17d69170c64ca2d2L,
	0xacca77bc2d549986L, 0x8a766ea0c0048193L, 0xbbb9c4e6c8dc7267L, 0x78702d8b2f37426L, 0xff11ee26c5507f57L,
	0x4b77e5f6c96b721fL, 0x27cc52791f0134c8L, 0xf6872aa510599eceL, 0xeb6877399d511f01L, 0x426c9f06cd7f2b9eL,
	0x4fa94ed19d3e01d6L, 0x3ffbf5595e325abbL, 0xfe543ec56c3adc8fL, 0x2093aa28ee54ba95L, 0x7aa8aec232638441L,
	0xbf6f8b045a6dfeb6L, 0x6a72f557df8503f6L, 0xee532835849974a7L, 0x184ebf7665c04da7L, 0x2b7f4a52e3898b51L,
	0x8a1c33c7ae5ca8d8L, 0xc5275ec4b0c63cc9L, 0x89fe93f9fd966d8eL, 0x882b1f1f947fac5fL, 0xcd2ff4e2b874a4b2L,
	0x3a2c69a2fc48f227L, 0x7af8c15ca55b30feL, 0x9273af29d074ab14L, 0xb849c5acc9974ecdL, 0xfcadb1e672c9a0ceL,
	0x385638a720701bb2L, 0xcb28bc3f93c742e5L, 0x8ab1123693c0febfL, 0x127dd4b1f7e55107L, 0xa1271598bfe203a5L,
	0x75956ef8835b2a69L, 0x16bb6759b804d25fL, 0x3d94a4baa0ade1abL, 0xeafb4cba0653fc6eL, 0x363d53c1a316c3ffL,
	0x875060309cadc94aL, 0xc4c2554135f5870aL, 0x517d5aa43bdae7e7L, 0x3a97f8114c2dfbb1L, 0x1ea296d67bd8d2eaL,
	0xf0493affa7c41c0L, 0x46d9e19052ac1087L, 0x84529e4b4ae6aceeL, 0xf9de603bf3a751bcL, 0x9f21dd4c1a24ad7aL,
	0xa945df6c6a3a5072L, 0x7801aa318d1c87a3L, 0x606e7ae92c36d0aL, 0xaf2f7e1a7cedc955L, 0xd71660c3519539fbL,
	0x3236c700c52284a5L, 0x570dab9bf9f78497L, 0x4fde0438cb17ce9fL, 0xe02ff0b55b9342a5L, 0x6a7630d1b7a52245L,
	0x80aa3dc478a94a5dL, 0x9355118ad5a9ba0L, 0x557180d7472a705cL, 0xfe31ec377125700bL, 0x26f3df1e4388a93dL,
	0x8d3c27a8f94c3d81L, 0xb8671186a4f593f6L, 0xd897a271d1f08ceL, 0xc1373a5b61962f62L, 0x50745c08be2d862aL,
	0xa5811964c8472c29L, 0x3df5f54c9fc82774L, 0xb14c125511c76ee2L, 0x9098f4656f815a93L, 0x73e9e8e3bdff3a88L,
	0x23e0737bf831aabbL, 0xfdf131057fcfe99dL, 0x784f9d568bf90512L, 0x63df2f7197051873L, 0x5dc78dd55a754094L,
	0x5abb4e141fa7e8c6L, 0xbd7d8a556d6062feL, 0x3a363ebd9e2cfd20L, 0xc4c55a16e4e3159bL, 0xf21a8dff31b8e92bL,
	0xc011e95af031c398L, 0x8ee71b210c8ee379L, 0x5347e3b632ffbf2fL, 0x3f8baf5b92bf4f40L, 0xa447e3bf61f76719L,
	0x186141f21cf6150cL, 0x507fe3d9c03e62eeL, 0xe5a5bbc395edbef3L, 0xdaa48dc759ce1bfbL, 0x7fdae33785c26a10L,
	0x583d4eb82cc44832L, 0x5ee3f34b78022f23L, 0x5324166fb71a04c0L, 0x92a457999c208fbL, 0x6c97e170c5bfab37L,
	0x106b10213997f878L, 0x8752752117adc054L, 0x15d7208cefe797eaL, 0x77c7c062bb9eec35L, 0x32c8de9c40e0940cL,
	0xb6f05080826fc14fL, 0xa78b482d7da0dae8L, 0x1c752552069ad3L, 0x35be864bd3ddf22eL, 0x210fb128e897d0bL,
	0x610b2c3405eb524L, 0x9b046d3f8f7b58bcL, 0x4c779e0876903c3dL, 0x53e4e9477c3e339bL, 0x2b3c8a2d4d9c3581L,
	0x2e663585f9f7db31L, 0xaa915e267496474dL, 0xdbc99cc570dcf72dL, 0x1727d55d63005e82L, 0x61644b66227c02d9L,
	0xb69b733ed528b2fbL, 0xaa906b1e799fc64cL, 0xa086c6f583659c50L, 0x51461495fb6b264eL, 0x48301753f305e8c7L,
	0xa05b2012d35e1bfbL, 0x29ca68b840140849L, 0x203a1b7bc0f69be4L, 0xc312351c060d9f1dL, 0x7e2eb7547ce70fb4L,
	0xeb18fe4e78d14e68L, 0xde788a308bb167f4L, 0x29afe89fa9c7238cL, 0xa871704ec312dc54L, 0x41db224f9aba36b7L,
	0xc01850eb8b79daf4L, 0xeb4921f09ca327bL, 0x67c2e9b92a6449d3L, 0x69c776ee6dc6432bL, 0xa7154f5f96804ea5L,
	0x550e146b7dacae1fL, 0x6a12b804bc91bf1bL, 0xc3e85461bfaa0ca9L, 0x6371a4c9f800a99cL, 0x1cabc45b5f8da3f3L,
	0xf8e783e89eef83deL, 0xbc53e6b9407273c3L, 0xb780d3f2083bc6ccL, 0x30da4ab0caa89ebdL, 0x6a31274db164a7c1L,
	0x826846c56130301L, 0xf41b5613077e2315L, 0xe0754514b8e9574aL, 0xd8f49e4310281c86L, 0x8590228a87441f7aL,
	0xc7114bd11f2c56c9L, 0x5c049b10ae527890L, 0xd993195f348d135cL, 0xe2306eeb4dffaf2dL, 0xbe7d771f5a3afefbL,
	0xab25c92bf478883eL, 0x1db1f24bd6f8358L, 0x5e31ea403dac5dL, 0x978a08c13a034e40L, 0x909bd3fc09cedb07L,
	0xa871f29d25931487L, 0xfd9cc0dad7bc4612L, 0x1539b5603783340L, 0xa5fae3d2cf8fda43L, 0x800f78a8ce8e18bcL,
	0xeb625eca240fe391L, 0x7e4f8cf31996c69aL, 0x8e89708f66fe91c2L, 0xa44677384afe9b7bL, 0xd5ab10478d6ae762L,
	0xd45d68a46d190e03L, 0xceb4b78ec600e6fbL, 0x5ed58b9069b487d0L, 0x52e2376f6cfe5475L, 0xe51ddea0832d5f72L,
	0xefa7d4fefa87970aL, 0x859fd0c24abfb0e8L, 0x7b41427b04cbeeaeL, 0x1446f162659d19baL, 0xf60d8a9204e23f01L,
	0x158d76b20a6ff2eL, 0x234effb0b5653989L, 0xc599e31bdd297c0eL, 0x3d4337925270ea37L, 0x5dea5ff6ff016692L,
	0x2d3a81a758a69e25L, 0xfaf08eb18f8e4bL, 0xb540cf102a2f7bc0L, 0x91dd7915ca509c40L, 0x8df04e5d3856f9f9L,
	0x8dc77b0032000c91L, 0x8669237a947aed5L, 0x74eb6e996639f848L, 0x7b3194aa1bf4d96aL, 0x38b327989a828e7dL,
	0x6fd218284a0964b8L, 0xee17a366290a1a07L, 0x80e65c2ad66d815fL, 0x6b845211074f6528L, 0x8813e8fce198d051L,
	0x362e40cfbca49656L, 0x3ac62d90b2afbd09L, 0x3aaf72190be84bdbL, 0x971dd8a601860909L, 0x18bebba7ca690531L,
	0x4e7d611afbd42714L, 0x5832b8f26a395e07L, 0x7377c8f16506b740L, 0xadb6e2c13d81acc8L, 0xcb22a0f46e9b740L,
	0x4e50d6d6a1c1b7a3L, 0x6a76a33fc218dafeL, 0x1ed41da1536ce322L, 0x586ab29a1fd48c60L, 0xbe0f71fbe04aeaecL,
	0xdae1571a23fae917L, 0xf23311da6cdb0ffL, 0x58a4a695bb032ce1L, 0x497a82682a67c21aL, 0x3d39a08645fb7d95L,
	0xf90ebdbbf2679ce1L, 0x484a85b63c8802d6L, 0x9e601fa5cad745aL, 0xb285893ab2b42bf8L, 0x81bc9ad4d45b21feL,
	0xfbb6baed220d058fL, 0x746029665929fe99L, 0x8f58f45c66384bb4L, 0x7935bf5a8e80a0acL, 0x1fbbd5df807f31deL,
	0x3206a8d4c85ad24aL, 0x296f755bbffc5aafL, }
lung = 0x88378e050bf043b4L
	\end{verbatim}}

 \subsection{Bad seed for the Well19937a}
 \tiny{
\begin{verbatim}
seed = { 0xc0541cc4, 0xab37d083, 0xb6fc37a, 0x1fbfd4f7, 0x7be7aa03, 0x2adda123, 0x798486a, 0x1dd1f7b7,
	0x297686df, 0xaed7e9e6, 0x394b50ff, 0xfc9767f8, 0xaf9a256b, 0x1ff43ac7, 0x5f720189, 0x232e9a24,
	0x699084cc, 0x4a8adc2, 0x93ceccf8, 0xe9b20d1, 0x4aa24890, 0x5dbf6264, 0xeb667014, 0x951df26f,
	0x9ff13450, 0xbcd06f9a, 0xe09e7cf5, 0x93fd5907, 0x1de2c6d0, 0x9068c876, 0xcec468d7, 0x68b48065,
	0x4a641145, 0xacc2bd32, 0x569200e8, 0xc04a709f, 0xc783924, 0xefdaa8b5, 0xcc84bce9, 0xa04e425e,
	0x910d5f58, 0x7fd365c4, 0xe5f93f70, 0x5f308f65, 0x1038bfeb, 0x416b61a3, 0x204867d7, 0xa85c5113,
	0x219457a8, 0x1ea7e3bc, 0x6d30742, 0x9ab602b3, 0x3e9abb1, 0x175a06d6, 0x5f09402, 0x93bb1830, 0x89f62013,
	0x987d013f, 0x337d257d, 0xee00c67b, 0x19b781f3, 0x977ef379, 0x6f901d2a, 0xda0ea75, 0x5336774d,
	0xa5418c03, 0xc8a3dbc, 0xb6b8275b, 0x8fb4059c, 0x56ca0d3d, 0xfbb7d386, 0x24ee9e0e, 0xa82552d1,
	0xc3fd647e, 0x1e405ade, 0x9bca4d1a, 0x3efa0811, 0xaedb503, 0xeb63fa60, 0xf258844a, 0xec3bc4d8,
	0xd123b06d, 0xf0fcb3e9, 0x126af477, 0x36c95046, 0x99a69493, 0x778f44c7, 0x4bd40dc1, 0xc339283d,
	0x7c920d82, 0x64c7fac4, 0x1d48acc4, 0x4803a8f9, 0x1f685f74, 0x4cb50562, 0x112eeac1, 0x55ff63b7,
	0x2e68517f, 0xe55dfb89, 0x8fe9cd84, 0xd56d399c, 0x1e81cb40, 0x50b3a0b3, 0xfffac543, 0xe3c9336d,
	0x3916deee, 0x151e0272, 0xb0d054ef, 0xf56167f4, 0x8d73ea06, 0x1e4bcb65, 0x9e10ae76, 0x2d543b84,
	0x89830bcf, 0x91a87f9f, 0x784ecd4e, 0x6b3f9834, 0x2a846ddf, 0x90d4acc2, 0x792cab7c, 0xc0e77ab3,
	0xfc4afb0d, 0xa64ea858, 0xc07fd5aa, 0x17395d36, 0xf0d3e73f, 0xd37fbc3a, 0x8980984a, 0x3e5a1bd3,
	0xeec10711, 0xb709cc08, 0x5384a33a, 0x819df9ce, 0xa033f244, 0xe3c700bb, 0x1f13679c, 0xcffac343,
	0x3eb1fe1b, 0x6ed78a30, 0x52027d21, 0x4c419896, 0x6f60f235, 0x3c6abac, 0x3246e346, 0x53b836f2,
	0xda03a8dc, 0x220a23b8, 0x74db4f52, 0xcf7e56f3, 0xb47f0949, 0x648affcf, 0x8ea280c1, 0x64d0071b,
	0x325c9a5f, 0x241366f2, 0x7e680e88, 0xe3a5ca68, 0x79453176, 0x56d744f5, 0x1e69aa63, 0xd66b1dca,
	0x299e580d, 0xc7a65801, 0xef948205, 0x67940480, 0x7cd16ec1, 0x4b203c8f, 0x8622eb56, 0x1a3eba21,
	0xb025c06a, 0x94b0c865, 0x364543dd, 0x84dda74e, 0xe55fc1b0, 0xf24015b3, 0x72567a7b, 0x80200378,
	0x661e0f4f, 0x8035ac86, 0x5ec704b9, 0x8b7e236d, 0xae20cca5, 0x7c6d784, 0xc79223b8, 0x273d8b80,
	0x356d44ba, 0xd469ac4e, 0xa300253c, 0xfbd0d15a, 0x182ba4dd, 0x583a3f38, 0x7cebcf17, 0xed18a530,
	0xb606c692, 0xdd2dece3, 0x9f9eabb, 0x6bd01a9c, 0xf1cf01ed, 0x347fca04, 0x7310f74f, 0x51603d78,
	0x474e8a04, 0xf2ec3806, 0x41b4723f, 0x9b4dbda2, 0xbbae52, 0x3976cc86, 0xe51820c, 0x10f518fc, 0xa563ed0d,
	0x54a23375, 0xc55ffe7c, 0x18100b8, 0x5ab4388b, 0x586761a4, 0xa2abcfea, 0x3e4e1d2c, 0x605ab89f,
	0xb10e6f94, 0x1f217537, 0x7054ed8f, 0x9f89b8b, 0x9cc7047d, 0x39962ba0, 0x1c8b79c4, 0x343b995f,
	0x4241b7de, 0xccf0b0a3, 0x7bef01e3, 0xe24af80e, 0xd39ad3eb, 0xab0181cf, 0x82913f77, 0x714646b8,
	0x2c4882db, 0xaa26c544, 0xfbee353a, 0x3c0f6b97, 0x645e51a0, 0xe97c5a23, 0x281d8a80, 0x47fe49f7,
	0x19084b74, 0x54253681, 0xabaf321a, 0x4fd6ce98, 0xfddd595a, 0xac19fe3c, 0x7dd5e43f, 0xaf8ff1c4,
	0xee2c77d9, 0xcdc63470, 0xe63b1318, 0x6eef9aed, 0x3c6c275a, 0x3a160556, 0xe7bdd27c, 0xe5629ccc,
	0x5ec06be4, 0xb715abac, 0xf1ec2528, 0xc6c1de0, 0x9893c7db, 0xf76c6779, 0x97bc68ff, 0x7d4636cd,
	0x2ff1ad03, 0x9ed0a1db, 0x5ddcde91, 0x817148f2, 0xf074fea3, 0xbe0aadaf, 0xc9f054ab, 0xd8683936,
	0xff0f6f94, 0xf6af9245, 0x7883f226, 0x2d7ff153, 0xd32583ca, 0x6cb1c030, 0x1223a0be, 0x2ef4b433,
	0x9aeb7b61, 0xfff23b29, 0xde7fcdad, 0x2b651f6, 0x6d3b7281, 0x184b0aec, 0xfc065efa, 0x939b8a8c,
	0x1a49430d, 0x4924c717, 0xbda1940a, 0x539d6e96, 0xdc6f7803, 0xc716d8f8, 0x24c6dfef, 0x64449248,
	0xf734c5ac, 0x7a42820b, 0xc63b9141, 0x63a64882, 0x838c6211, 0x5b8d488d, 0x3524a121, 0xb63eb35c,
	0xe8117c0a, 0x68ee3556, 0x7eb37b94, 0x5ddac590, 0x1a00b9a7, 0x3ef0e2a0, 0x22f09b06, 0xb90ff34e,
	0xff0d7aa5, 0x3dc0a7a0, 0x9fb70a22, 0xf23894d, 0x2dec32d6, 0x17ceb753, 0x75d711f3, 0x6482cf8e,
	0x39336eb5, 0x6ef95556, 0xc87c30c6, 0xf9b99870, 0x45e352cb, 0xe23598f9, 0xea74a0ac, 0xced93d48,
	0x76bc9106, 0x61e65a65, 0x2c793020, 0xf0ff6f22, 0x1d7e5a84, 0x840b85fe, 0x7d22f3f4, 0x3dbacba0,
	0xdaf6eb9, 0x521ce21c, 0x40ecaab4, 0xbb33041b, 0xedc07d5, 0x7ee249fa, 0x34454402, 0x711c89ee,
	0xe75ad56b, 0xbd609f1a, 0xcd371052, 0xbc23f095, 0xdde9c552, 0x8301cca9, 0x576754a7, 0x2097e4a7,
	0xe75c716f, 0x74578e4c, 0x8b8064b4, 0xa2437e2b, 0x3a9def3, 0x942057c4, 0x5931363f, 0xa73bbe28,
	0x186fe95d, 0x6bb85632, 0x6df0b3c, 0x76692f9, 0x9ea27b47, 0x347c0f6d, 0x82283e6b, 0x6d894175,
	0xb418f0ae, 0xfbda128a, 0xdc906372, 0x94300dd0, 0xc6b7f71d, 0x148e9e4a, 0x22700dbf, 0x5b5fd968,
	0xb1f930af, 0xdadc78c4, 0xe64609ea, 0xf692ff99, 0xdd0dcaa6, 0xa78ad6a8, 0x6efb8200, 0xfa9f0c91,
	0x6b2c0f66, 0xcc11eb9, 0x8d494af3, 0x949d08dc, 0xf525c8fc, 0x426adcde, 0x11e2137a, 0xea08c44a,
	0x8a36d4ef, 0xce779ca4, 0xa4ca9bcb, 0x8063de48, 0x417c9bb6, 0xec1a5e8e, 0xfbb7b631, 0x3f7a4e04,
	0x625f41b6, 0x60b004e9, 0xe530cc9, 0x92bbbb24, 0xc573ce44, 0x6d7b1579, 0x8c5aa5d3, 0x3074eaeb,
	0x6685214, 0xb14fedf5, 0x5409bf8c, 0x98b4dd15, 0xee872716, 0x153fa012, 0x75b2cd43, 0xdf224178,
	0x8497066c, 0x2bd58211, 0xaf70c531, 0xc2f2a33a, 0xeaec04ec, 0x1ec1f47f, 0x345bdfde, 0xd93e5c7b,
	0x4e466a9, 0x5b02aa60, 0xf638de72, 0x73c0664, 0xd6ceb6b1, 0xf1141dca, 0x42bf744c, 0xfcbbde18,
	0x9d240cea, 0xc611523b, 0x63185cfa, 0x4642bfb3, 0x95c1f4fe, 0xec6d2818, 0xb7a4def, 0x727f8418,
	0x3cda2129, 0xe147ffbe, 0xcca78d21, 0x4aa50931, 0x15031031, 0x9aff12ad, 0xc0e98d0, 0xf68e013e,
	0x2eb22894, 0x218ea92a, 0x311fcbd6, 0xdc6dd8b6, 0xdf22949d, 0xafb228e3, 0x130b5bd4, 0xa0e9285f,
	0x81ca17ef, 0x5cd0c58e, 0x558480c5, 0x163696eb, 0x7f7ab974, 0x27f0ded0, 0x44947fdf, 0xbad30dbc,
	0xdc0ae75b, 0xf924e835, 0x29632fe1, 0xbaeaeb0a, 0xab89e1f, 0x5e0f07d7, 0x9b1807e2, 0x7623bd3e,
	0x50f43bb2, 0xb1b5697d, 0xa3427405, 0x53bb9236, 0x3b13778a, 0x269f862b, 0x247aa08b, 0x27b5d209,
	0xa3bf5f5f, 0x75e1e965, 0x130f70e1, 0x828638ed, 0x6ecfbc82, 0x495cba91, 0xac53da32, 0xa087293,
	0x6d15e358, 0x3c45c3b7, 0xbb6f605, 0xac2c7586, 0x69f2f9d4, 0x95843ecb, 0xc7619464, 0x592d6845,
	0xe964a70d, 0x14e04d4d, 0xe27b66c9, 0x6c938eb8, 0xfeaf5f29, 0x17c4fcb8, 0x9ea0a4e9, 0x668fd952,
	0x80cdefd5, 0x6e83a47d, 0x1510ee2c, 0xe655cfe1, 0x98a2fb6e, 0xeaec016c, 0x62859741, 0xf7f1c678,
	0xce4d049b, 0xc02db0b2, 0x6e726be8, 0x58089b89, 0xc3a9eb5d, 0xf2d5a912, 0x4e92cae3, 0xad6decf8,
	0x2de3c1e3, 0x950cea2b, 0xdcd56e67, 0xce9a19d, 0x982ea67d, 0xe68ab9fa, 0x93ac8b98, 0x30b8fcde,
	0x42e106b6, 0x46a4059, 0x2d981267, 0xec4c02b4, 0xf4af6048, 0x6972782b, 0xbefb1f4a, 0xb0bb565b,
	0x2eca8d5e, 0xe3250ba9, 0xc5b4fa98, 0xf19b5c4, 0xa5e3d150, 0x784b672d, 0x3c5fe64, 0xab4e060e,
	0xc2118c74, 0x3119e315, 0xc1579db4, 0xfd608c3e, 0xf886117b, 0xd8c6c3d1, 0x12ce0011, 0x2bae2a49,
	0x5c3b1813, 0x51a4eb56, 0xe015b1a5, 0x36f794d6, 0x7d8985f2, 0x2dcf9571, 0x35e48ba, 0x5e454bbf,
	0xe21f8ea1, 0xbe45509e, 0xc478479e, 0xeb0b46d, 0x64c2cb2c, 0xee1a7563, 0x2211ccc3, 0x65191c99,
	0x3abfa924, 0x51ac52d9, 0x311f465a, 0xd9883ef4, 0x757e0867, 0x725e8a93, 0x46db23e7, 0x5e277622,
	0x58b93c88, 0x29f9c42f, 0x45855d3c, 0x7ea87197, 0xf1117e94, 0x6f4b2869, 0xe5d5c993, 0x823694a5,
	0x81d2a628, 0x23336275, 0xef5d5c2a, 0x6c17fc14, 0x987b0bce, 0x4e915b57, 0xcdcbc270, 0x4b5f67b5,
	0x79cedd61, 0x9d183ddb, 0xd55256a6, 0xd093815b, 0x232e9676, 0xb66bae1, 0xae33e73f, 0x2c371d4d,
	0xd6a3ba98, 0x3df264d5, 0xfd799bae, 0x4edb96e9, 0x63b8db44, 0xbfe65c1f, 0xec967faa, 0xfb848f6a,
	0xaf72e86b, 0xbd23809c, 0xa367205d, 0xafd4f468, 0x92b66d15, 0xe8696889, 0x9757015, 0x87874cbf,
	0x5c20a641, 0xc8b0387f, 0x4ea381d4, 0xccf6bcf7, 0x1aea51b, 0x8b12819d }
\end{verbatim}
}

	\section{Eigenvalues of some Mersenne-twister variants}\label{appendix:eigenvalues}
\begin{figure}[h]
	\centering{
		\subfigure[\label{fig:mt19937x64}MT19937x64ID1]{	\includegraphics[width=.4\textwidth]{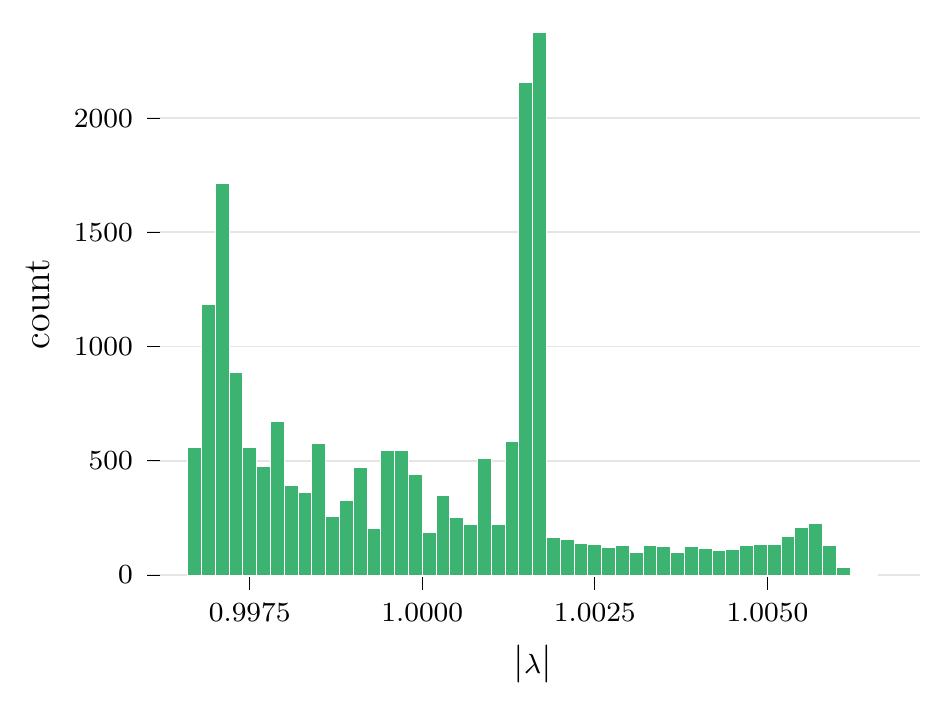}
			\hfill
			\includegraphics[width=.4\textwidth]{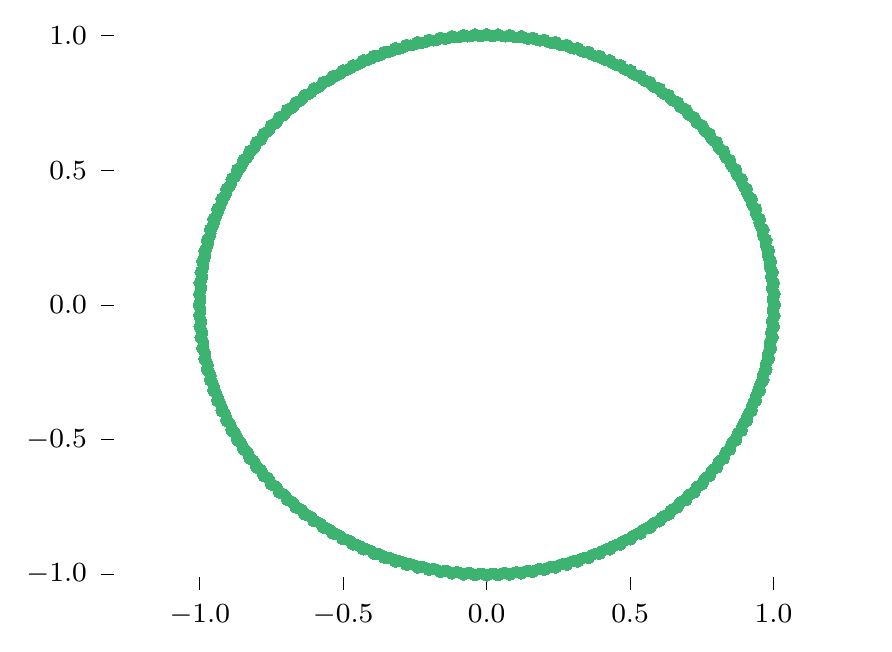} 
		}
		\vskip\baselineskip
		\subfigure[\label{fig:mt19937x64ID3}MT19937x64ID3]{	\includegraphics[width=.4\textwidth]{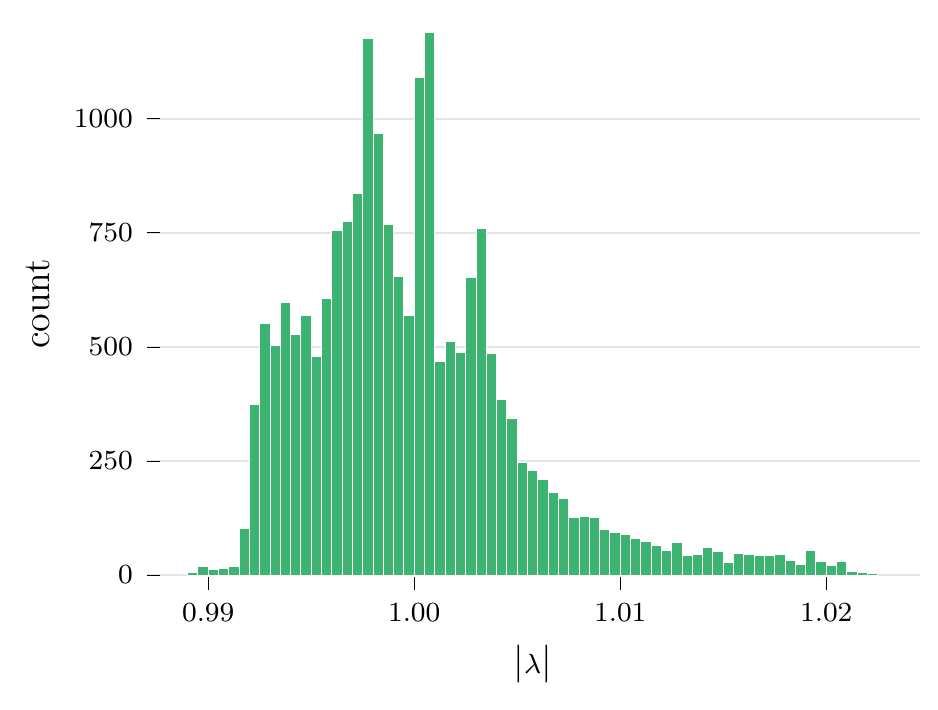}
			\hfill
			\includegraphics[width=.4\textwidth]{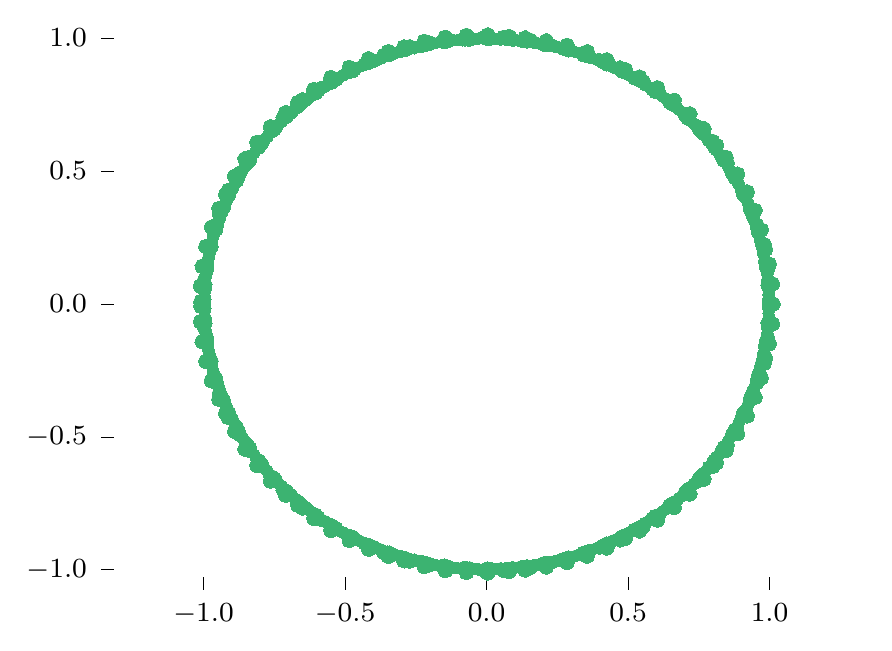} 
		}
		\vskip\baselineskip
		\subfigure[\label{fig:Well19937a}Well19937a]{	\includegraphics[width=.4\textwidth]{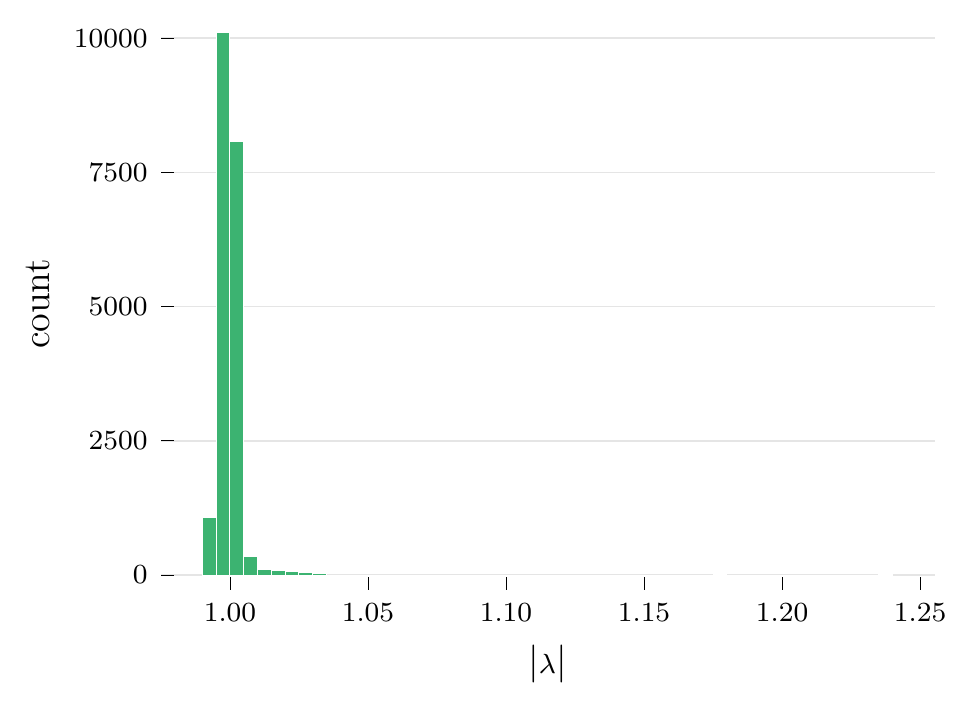}
			\hfill
			\includegraphics[width=.4\textwidth]{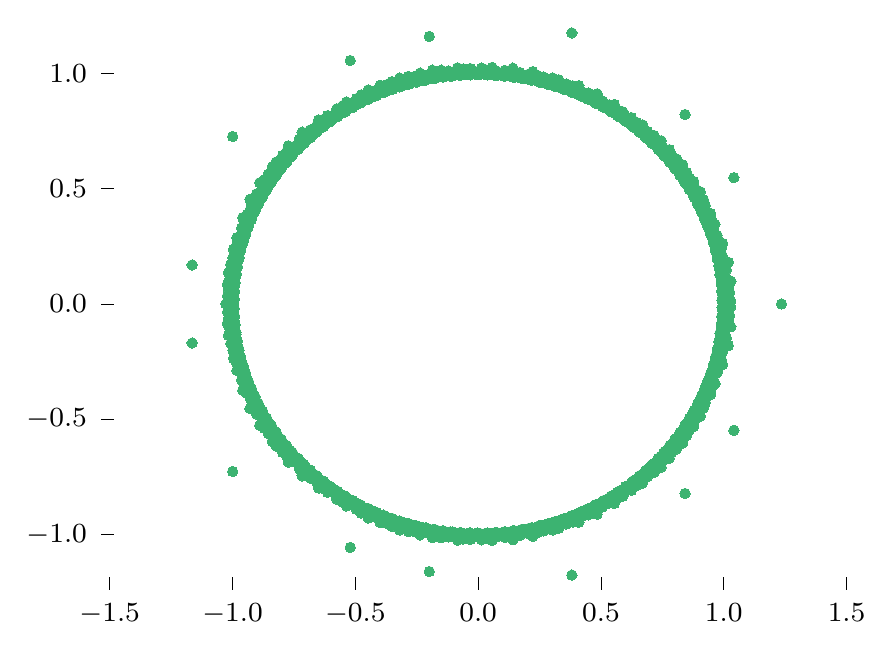} 
		}
		\vskip\baselineskip
		\subfigure[\label{fig:Melg19937a}Melg19937]{	\includegraphics[width=.4\textwidth]{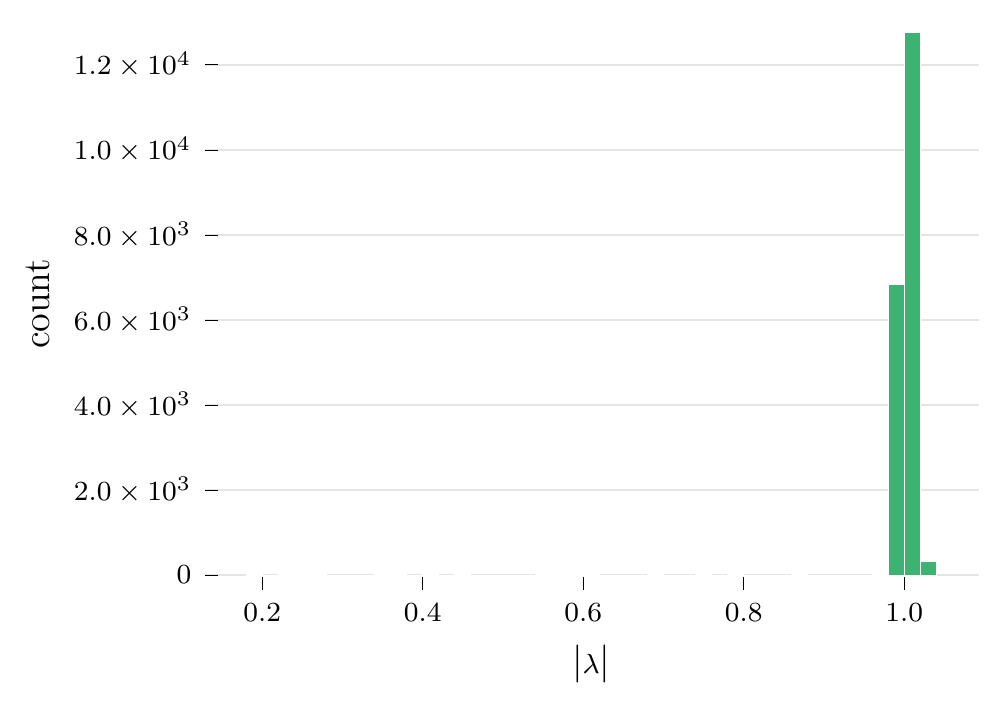}
			\hfill
			\includegraphics[width=.4\textwidth]{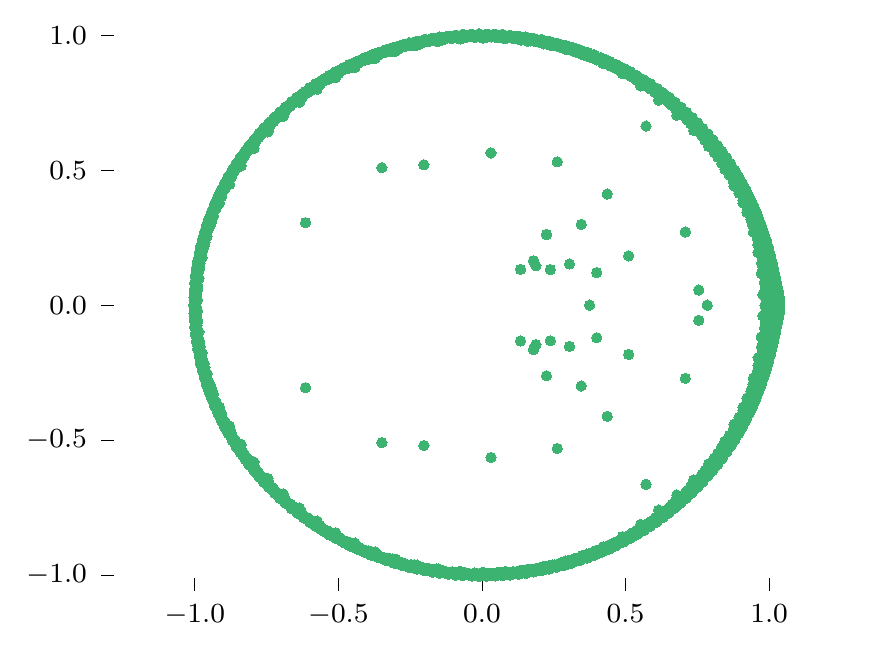} 
		}
		\caption{Eigenvalues of the transition matrix of generators with a period of $2^19937-1$: histogram (left) and complex plane representation (right).\label{fig:eig19937}}}
\end{figure}
\begin{figure}[h]
	\centering{
		\subfigure[\label{fig:well1024a_hist64}Well1024]{
			\includegraphics[width=.45\textwidth]{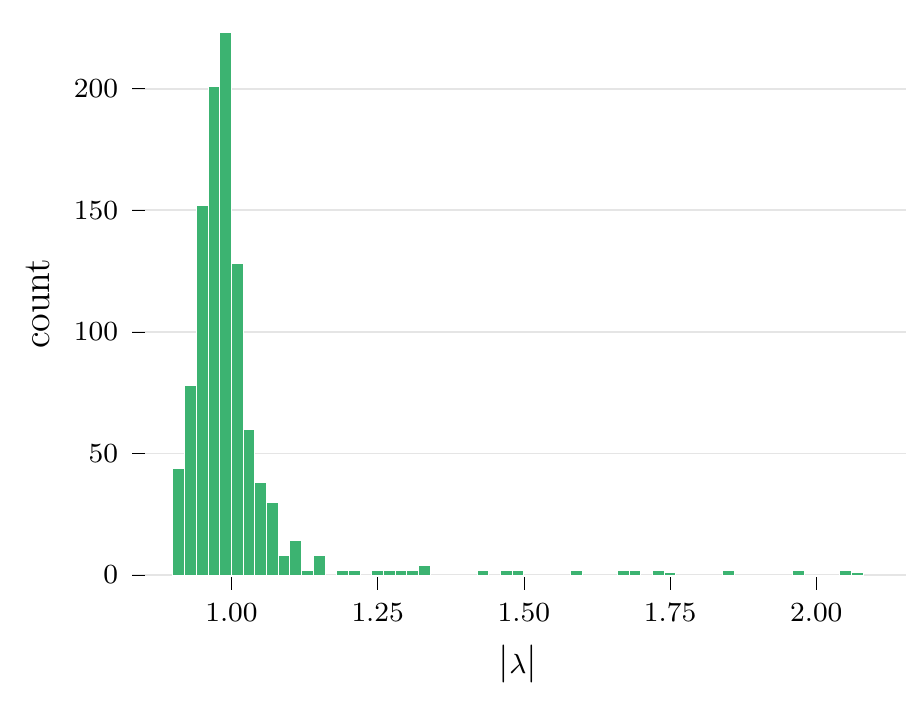}
			\includegraphics[width=.45\textwidth]{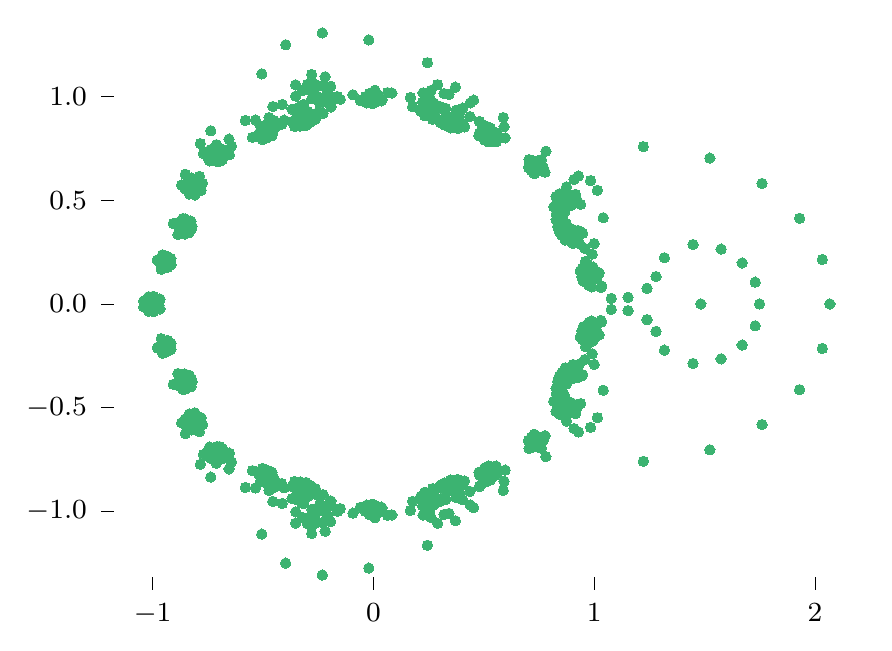}
		} 
		\subfigure[\label{fig:well607b_hist64}Well607b]{
			\includegraphics[width=.45\textwidth]{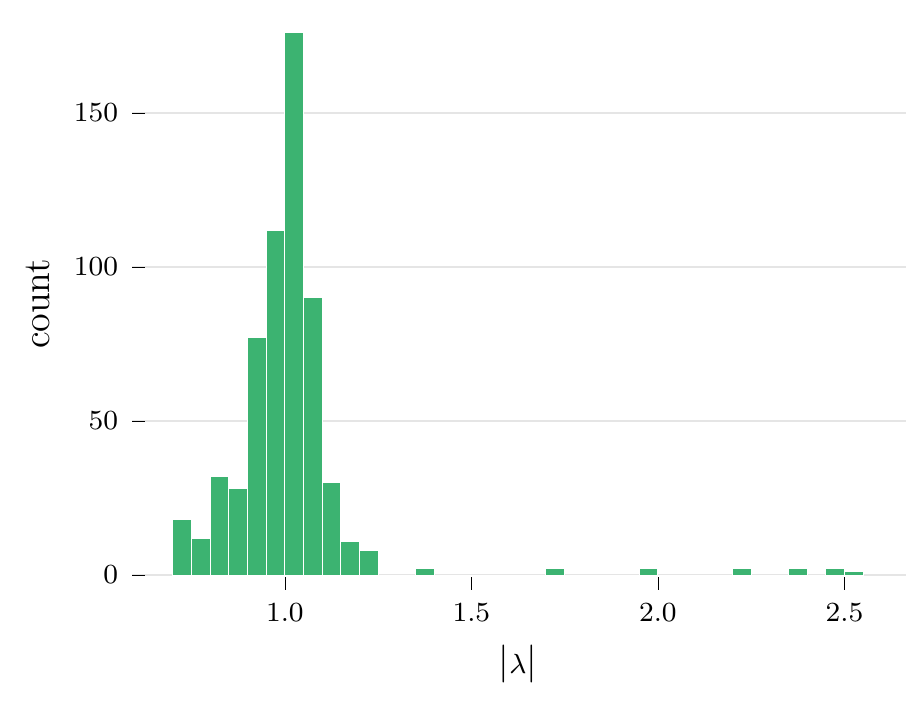}
			\includegraphics[width=.45\textwidth]{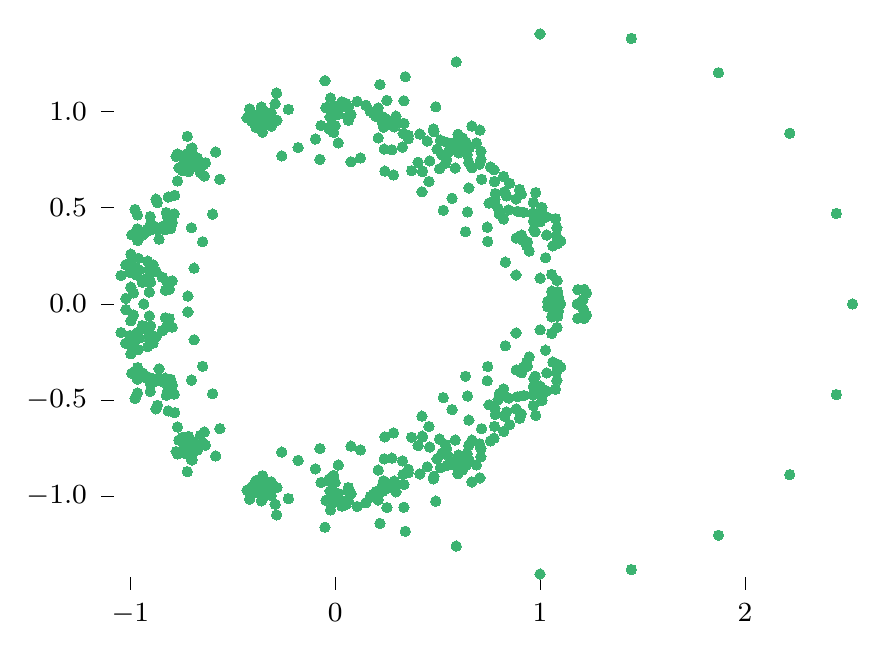} 
		}
		\subfigure[\label{fig:melg607b_hist64}Melg607]{
			\includegraphics[width=.45\textwidth]{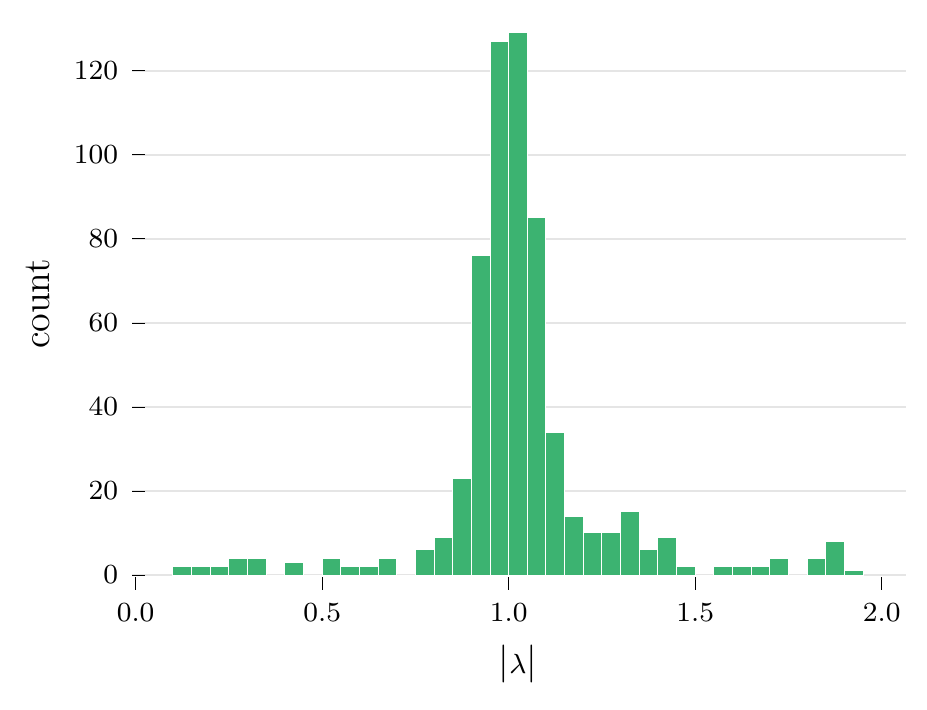}
			\includegraphics[width=.45\textwidth]{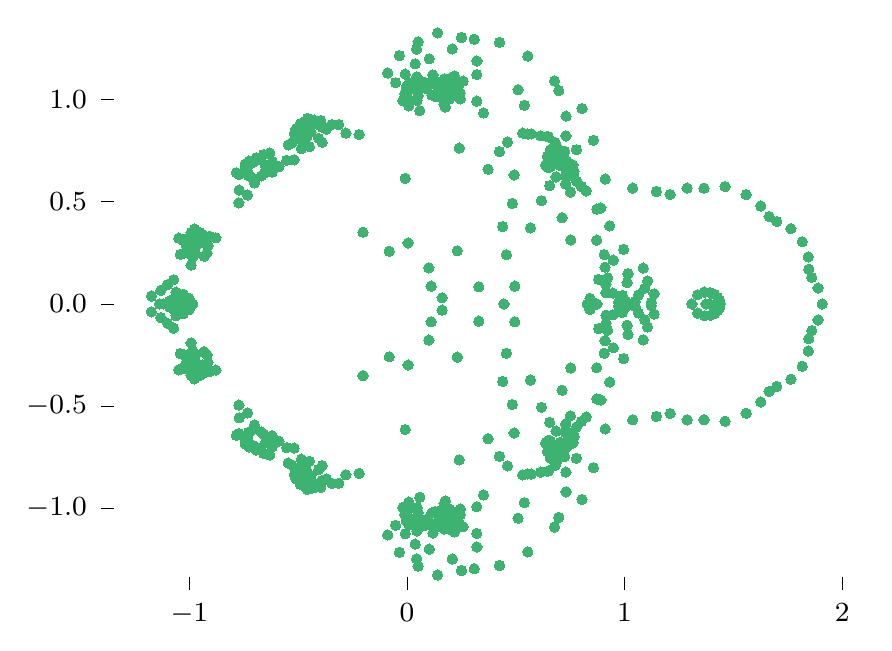} 
		}
		\caption{Eigenvalues of the transition matrix of generators with a smaller period: histogram (left) and complex plane representation (right).}}
\end{figure}


\end{document}